\documentclass[aps,prl,twocolumn,groupedaddress]{revtex4-2}
\usepackage{graphicx}
\usepackage{amsmath}

\begin{document}


\title{Hybrid roles of adaptation and optimization in formation of vascular network}


\author{Yawei Wang}
\email[]{wangyawei@buaa.edu.cn}
\author{Zilu Qin}
\author{Yubo Fan}
\email{yubofan@buaa.edu.cn}
\affiliation{Key Laboratory for Biomechanics and Mechanobiology of Ministry of Education, Beijing Advanced Innovation Centre for Biomedical Engineering, School of Biological Science and Medical Engineering, Beihang University, Beijing 100083, China}


\date{\today}

\begin{abstract}
It was hypothesized that the structures of biological transport networks are the result of either energy consumption or adaptation dynamics. Although approaches based on these hypotheses can produce optimal network and form loop structures, we found that neither possesses complete ability to generate complex networks that resemble vascular network in living organisms, which motivated us to propose a hybrid approach. This approach can replicate the path dependency phenomenon of main branches and produce an optimal network that resembles the real vascular network. We further show that there is a clear transition in the structural pattern of the vascular network, shifting from `chive-like' to dendritic configuration after a period of sequenced adaptation and optimization.
\end{abstract}


\maketitle


\section{Background}
The vascular networks play crucial roles in various biological functions and processes within living bodies, and have attracted widespread interest in several scientific fields \cite{O’Connor2022,Kinstlinger2020,Banavar2000,Schreiner1993}. There are some intrinsic characteristics about the vascular networks in living bodies, including: (1) large span of vascular diameter ranging from a few micrometers to several centimeters, (2) phase transitions in vascular structures at different geometric scales, and (3) path dependencies on tissues and interfaces of main branches in vascular network. Among these intrinsic characteristics, theoretical aspects of the topological phase transitions in vascular networks attracted a lot of interest \cite{Hu2013,Corson2010,Katifori2010}. The design and fabrication of vascular networks have also drawn many attentions \cite{Kinstlinger2020,O’Connor2022}. As concluded in a review paper that, `Our increasingly detailed knowledge of vascular network architecture and composition has made it clear that replicating the structural and biological complexities of vascular networks in engineered organs and tissues may require the integration of diverse technologies and fields of expertise' \cite{O’Connor2022}. It is still a great challenge to fabricate functional vascular networks in the present and the foreseeable future, and theoretical tools, capable of generating vascular networks that resemble the actual vascular network in living body, may provide key assistance \cite{Katifori2010}.

Biological transport networks have been studied in the optimization framework and the adaption dynamics framework in past two decades. In the optimization framework, the energy consumption within the networks is minimized under the constraint of constant total material cost \cite{Katifori2010,Corson2010,Durand2007,Banavar2000}. Since total material available to build the network is limited, the optimal results of this framework is global. In this framework, the phase transition of topological structures of the biological transport networks are caused by an exponent $\gamma $, because it is assumed that the material cost for an edge of the network is proportional to a power law $\kappa^\gamma$, where $\kappa$ is the conductance of the edge. The studies using this framework have shown that the optimal network exhibit a phase transition at $\gamma=1$, with a uniform sheet for $\gamma>1$ and a loopless tree for $\gamma<1$ \cite{Katifori2010,Durand2007}. In the adaption dynamics framework, a diameter-adaption model containing the real and constant optimal wall shear stress was proposed, including the effects of a driving stimulus of blood flow and the intrinsic decreasing stimulus \cite{Hu2013}. This adaption model can be naturally generalized to include fluctuation in flow distributions through defining an open probability of sinks, which help exhibit a potential phase transition for topological structures of the vascular network \cite{Hu2013}.

Since the studies in the optimization framework and the adaption dynamics framework exhibit their ability on simple two dimensional spaces \cite{Hu2013,Corson2010,Katifori2010,Durand2007,Banavar2000}, it is reasonable to propose the question that, whether these frameworks are capable to generate vascular network resemble that in living body. As a widely circulated saying goes, `more is different', especially for complex nonlinear systems \cite{Anderson1972}. Real vascular networks in living bodies usually have irregular three dimensional geometrical spaces and include different types of tissues, such as bone and muscles. Vascular networks in living bodies typically exhibit a characteristic path dependency on tissues and interfaces. For example, main branches of coronary arteries usually locate near the surface of the myocardium, for avoiding forced collapse during myocardium contraction \cite{Chiu2012}. Similar path dependency phenomenon appear in many living organisms, such as brain and skeletal muscles \cite{Gautier2000,DELONG1973}. In this letter, we test the abilities of the optimization framework and the adaption dynamics framework on generating vascular network in a test space from a rat lower limb firstly, and then propose a hybrid approach based on these tests, and explore the intrinsic features of the generated vascular network finally.

\section{basic physics and the test case}

Tubes form the edges of many vascular networks in living body, and the inner flow is usually laminar (except in some large blood vessels \cite{Pries2001}). The laminar flow in the blood vessels can be described by the Hagen-Poiseuille's law, $ Q = \kappa \Delta P$ and $ \kappa = \pi D^4/128\eta L$, where $Q$ is the flow rate, $\kappa$ is the conductance, $\Delta P$ is the pressure drop, $D$ is the vessel diameter, $L$ is the vessel length, and $\eta$ is the viscosity of the blood. The metabolic flow regulation processes are capable of maintaining a relative steady mean flow rate in capillaries for blood vasculature \cite{Hu2013,Delashaw1988}. This provides the sinks for the arterial network and the sources for the venous network. Once the conductance for all segments and the flow sources and sinks are known, by analogy to an electric circuit one can calculate the flow distribution using Kirchoff's law, $\Sigma_j (P_j-P_i) \kappa_{ij} = s_i$, where $P$ is the fluid pressure, $i$ and $j$ are the indices of the two end nodes of edge $ij$, $\kappa_{ij}$ is the conductance of the edge $ij$, and $s_i$ is the given strength of flow sink (or source) at node $i$. For nodes not connected by an edge, the conductance between them is set to 0.

In the optimization framework, the total energy consumption contains two terms, i.e., the energy dissipation caused by inner flow and the metabolic requirements \cite{Corson2010}. To satisfy the principle of the minimum energy consumption, the conductance $\kappa_{ij}$ of each edge needs to satisfy the following formula (adapted from \cite{Corson2010} with $\gamma = 1/2$ for blood vessel networks), 

\begin{equation} \label{eq_opt}
    \kappa_{ij} = \frac{V^2 Q_{ij}^{4/3}}{8 \pi \mu l_{ij} \sum_{(k,l)} (l_{kl})^2 Q_{kl}^{4/3}}
\end{equation}

\noindent where $V$ is the total volume of the vascular network, $k$ and $l$ are the indices of the two end nodes of edge $kl$, and the $\sum_{k,l}$ in Eq.(\ref{eq_opt}) represents iteration over all the edges in the vascular network.

Natural vascular networks are usually in gradually adaptation for satisfying metabolic needs of biological tissues \cite{Chen2012}. Among the factors that will stimulate vascular adaption, wall shear stress caused by blood flow is a critical one \cite{Hu2013,Chen2012}. In this work, we use the adaptation model proposed by Hu et al. \cite{Hu2013} and set $\gamma=1/2$ for blood vessel network, 

\begin{equation} \label{eq_adap}
    \frac{d \kappa_{ij}}{dt} = c (\frac{Q_{ij}^2}{\kappa_{ij}^{3/2}} - c_0 l_{ij}^{3/2}) \kappa_{ij}
\end{equation}

\noindent where $c$ is a constant, whose value is positive and represents the adaption ability of the vascular diameter, $c_0$ is a metabolic coefficient, the term $Q_{ij}^2 /\kappa_{ij}^{3/2}$ is the driving stimulus of blood flow corresponding to the energy dissipation term in Eq.\ref{eq_opt}, the term $c_0 l_{ij}^{3/2}$ represent the intrinsic decreasing stimulus \cite{Hu2013}.

\begin{figure}
    \includegraphics{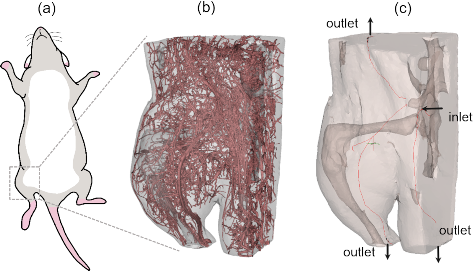}
    \caption{\label{Fig1} A geometric space carved out from a rat's lower limb. Three dimensional geometric model of this space is reconstructed from microCT images of the rat, and it is spatially discretized to form a fundamental mesh for the tests through using tetrahedral mesh with basic mesh size of 1 mm. All edges in the fundamental mesh are possible edges of the network. (a) Location of the geometric space. (b) Reconstructed blood vessels from microCT angiography images of  the rat. (c) Boundary definitions for the tests including one blood flow inlet and three outlets.}
\end{figure}

A geometric space is carved out from the low limb of a rat and used to test the performances of the optimization framework and the adaptation dynamics framework, as shown in Fig.\ref{Fig1}(a). This test space is defined by an outer boundary consisting skins and cross-sections as the outer boundary, an inner boundary formed by bone surfaces, and an inside zone. Simulated vascular network will be generated in the inner zone, base on an assumption that the inner zone is filled with uniform muscle tissues. The blood vessel network within the same zone of the rat is reconstructed from the angiography images and used as a reference, as depicted in Fig.\ref{Fig1}(b). One blood flow inlet at the location on the cross-section of the common iliac artery, and three main blood flow outlets are defined in the following tests, while small blood vessels spanning the cross section are neglected, as shown in Fig.\ref{Fig1}(c). This test space contains skeletal muscles that interface with the bones, leading to path dependency of the main blood vessel branches.

\begin{figure}
    \includegraphics{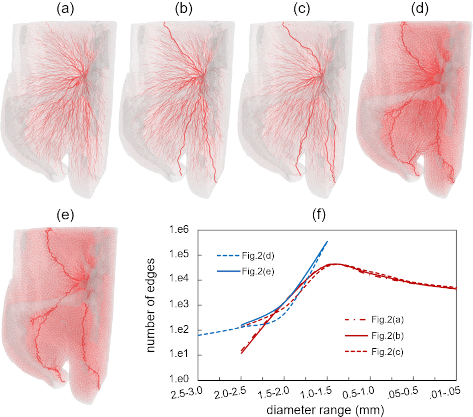}
    \caption{\label{Fig2} Typical vascular networks generated by using the optimization framework and the adaption dynamics framework respectively. The line width is approximately proportional to the value of edge conductance $\kappa$, which is proportional to $D^4$ and can better reflect the differences among edges. (a) The optimization framework with $ s_i=0.5$ and $Q_{\text{out}}/s_i=300$. (b) The optimization framework with $ s_i=0.5$ and $Q_{\text{out}}/s_i=3000$. (c) The optimization framework with $ s_i=0.5$ and $Q_{\text{out}}/s_i=30000$. (d) The adaptation dynamics framework with $c_{\text{s}}=0.5$ and $c_{\text{main}}=2.0$. (e) The adaptation dynamics framework with $c_{\text{s}}=0.2$ and $c_{\text{main}}=10.0$. (f) Variations of edge numbers with diameter ranges. }
\end{figure}

Optimal vascular network within the test space by using Eq.\ref{eq_opt} from the optimization framework is depicted in Fig.\ref{Fig2}(a). Generally, the structure of this simulated vascular network is `chive-like', whose main branches originate from nearly one point close to the blood flow inlet. This simulated structure is obviously different from that in Fig.\ref{Fig1}(b), as the real blood vessel have obvious main branches depicted by thin red lines in Fig.\ref{Fig1}(c). If one want to have fewer main branches in the simulated vascular network, a much higher mass flow ratio between the outlets ($Q_\text{out}$) and the given strength of flow sink $s_i$ is needed, as shown in Fig.\ref{Fig2}(b) and especially Fig.\ref{Fig2}(c). Meanwhile, the simulated main branches are nearly straight, running from the inlet to the outlets. This indicate that Eq.\ref{eq_opt} from the optimization framework do not have the ability to generate the main branches resemble to real blood vessel network. Distinct advantages of the approach based on Eq.\ref{eq_opt} include: (1) high calculation efficiency; (2) easy to control the total volume of the vascular network through adjusting the value of the parameter $V$. The second advantage can help control the pressure drop between the inlet and the sinks, which has important physiological meanings. 

In contrast, Eq.\ref{eq_adap} from the adaption dynamics framework can help generate vascular network that has main branches similar to those in Fig.\ref{Fig1}(b), as shown in Fig.\ref{Fig2}(d) and Fig.\ref{Fig2}(e). This is achieved by defining different values of the parameter $c$ in Eq.\ref{eq_adap} for main branches ($c_{\text{main}}$) and other vessels ($c_{\text{s}}$), where the subscript `main' denote the edges near the location of main branches of the vascular network and `s'  denote the other edges. Compared with the optimization framework, it is an important advantage of the adaption dynamics framework, to have the ability to generate main branches resemble real blood vessel network. Meanwhile, the other edges that are not near the location of the main branches are difficult to exhibit significant difference in their diameters after executing a large number of steps of adaptation (about 10000 steps with step size of 0.2 s in a test case). This may indicate that the approach based on Eq.\ref{eq_adap} from the adaption dynamics framework is inefficient to achieve completely convergent results.

Some intrinsic characteristics of typical vascular networks generated by using the optimization framework and the adaptation dynamics framework, are depicted in Fig.\ref{Fig2}(f). The maximal diameter does not exceed 3.0 mm, which is in reasonable agreement with reconstructed blood vessel in Fig.\ref{Fig1}(b). The numbers of edges with diameters ranging from about 1.5 to 2.0 mm generated by using these two approaches is highly consistent (about $10^3-10^4$). The approach in the adaptation dynamics generate more edges with diameters greater than 2.0 mm (several hundred) than those generated by using the optimization framework (dozens). This is reasonable when considering the total volume $V$ of the vascular network can not controlled. The most important difference between vascular networks generated by these two approaches appear in the numbers of edges with diameters lower than 1.0 mm, as depicted in Fig.\ref{Fig2}(f). The dynamics adaptation approach almost never generates edges smaller than 1.0 mm in diameter in the tests, while the numbers of edges with diameters in the same ranges decrease as edges diameters decrease from 1.0 mm. The reasons may include: (1) $c_0$ is set to a low value for avoiding negative values of edge conductances in the tests using the dynamics adaptation approach, which may lead to an insufficient decreasing stimulus; (2) the definition of sinks in the fundamental meshes in both framework may cause a `subgrid' phenomena, i.e., nodes with large edges passing through no longer require small edges to transport fluid, while small vessels like arterioles and capillaries are essential in real vascular networks \cite{Mohrman2006}. Therefore, the question of whether a simulated vascular network within a certain diameter range is reasonable depends on the basic size of the fundamental mesh to avoid the impact of `subgrid' phenomena. The adaptation dynamics approach may have the capability to generate vascular networks of full-scale, but the current form of Eq.\ref{eq_adap} with constant values for the parameters $c$ and $c_0$ will result in a very low algorithmic stability and efficiency.

\section{Hybrid roles in network formation}

Motivated by the performances of the optimization framework and the adaption dynamics framework, we propose a hybrid approach as following,

\begin{equation} \label{eq_hybrid}
    \kappa_{ij} = \delta \kappa_{ij}^{\text{(D)}} + (1-\delta) \kappa_{ij}^{\text{(O)}}, \quad \delta =
    \begin{cases}
        1& t\leq t_0\\
        0& t> t_0
    \end{cases}
\end{equation}

\noindent where the superscript `(D)' means the adaption dynamics approach and `(O)' means the optimization approach, $\delta$ is an operator for making a selection, $t$ is physical time and $t_0$ is a feature time for switching from adaption dynamics approach to the optimization approach. After switching to the optimization approach, the physical time $t$ lose its physical meaning and equivalent to the number of iterative steps. The term $\kappa_{ij}^{\text{(D)}}$ can be calculate by using Eq.\ref{eq_adap}, and the term $\kappa_{ij}^{\text{(O)}}$ can be calculate by using Eq.\ref{eq_opt}.

\begin{figure}
    \includegraphics{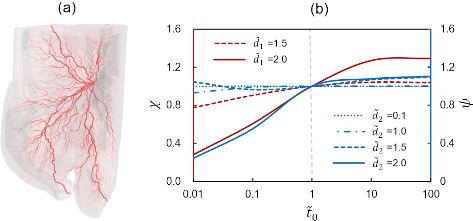}
    \caption{\label{Fig3} Typical vascular network generated by using the hybrid framework and its structural characteristics. (a) A formation of the simulated vascular network, and the line width is also approximately proportional to the value of $\kappa$ of edges. (b) Variations of the parameters $\chi$ and $\psi$ with the dimensionless time $\tilde{t}_0$, and the unit for $\tilde{d}_1$ and $\tilde{d}_2$ is millimeter.}
\end{figure}

A typical vascular network generated by using the hybrid approach is depicted in Fig.\ref{Fig3}(a). Different from the `chive-like' tree structures in Fig.\ref{Fig2}(a)~(c), main branches of the simulated vascular network in Fig.\ref{Fig3}(a) nearly uniformly originate from their parent branches along the flow direction. It is interesting and important to define indexes to distinguish the differences in these tree structures. According to the fundamental differences in their structural patterns mentioned above, we define two indexes as follows,

\begin{equation} \label{eq_disOri}
    \chi= \frac{\sum_{i = 1}^{N(\tilde{d}_1)} L_i}{N(\tilde{d}_1)};\quad \psi = \frac{N(\tilde{d}_2)}{N(\tilde{d}_2,\tilde{t}_0=1)} 
\end{equation}

\noindent where $\tilde{d}_1$ and $\tilde{d}_2$ are featured diameters, $N(\tilde{d}_1)$ is the number of branching points for edges whose diameter are greater than $\tilde{d}_1$, $N(\tilde{d}_2)$ is the number of edges whose diameter are greater than $\tilde{d}_2$, $N(\tilde{d}_2,\tilde{t}_0=1)$ is the number of edges whose diameter are greater than $\tilde{d}_2$ for results of $\tilde{t}_0=1$, $\tilde{t}_0$ is defined as $\tilde{t}_0=t_0/t_{\text{c}}$ (where $t_{\text{c}}$ is a character time when the energy dissipation term $ E_{\text{D}} = Q_{ij}^2 /\kappa_{ij}^{3/2} $ is convergent, i.e., $dE_{\text{D}}/dt < \varepsilon  $, and $\varepsilon$ is a small number). According to their definitions in Eq.\ref{eq_disOri}, $\chi$ can reflect the spatial distribution characteristics of the bifurcation points of the main branches in the vascular network, and $\psi$ can reflect the adaption process of the edge diameters across different size ranges.

Variations of the parameters $\chi$ and $\psi$ with the dimensionless time $\tilde{t}_0$ are depicted in Fig.\ref{Fig3}(b). The parameter $\chi$ for edges with a diameter greater than 2.0 mm continues to increase until $\tilde{t}_0$ reaches approximately 10.0, which means the average distance among the bifurcation points of these main branches expands until it reaches and stabilizes at an maximal value, and the increase will exceed four times the original amount. The parameter $\chi$ for edges with a diameter greater than 1.5 mm shows a similar variation with $\tilde{t}_0$ but with a lower slope. This variation of the parameter $\chi$ indicates that the structural pattern is gradually transitioning from `chive-like' to dendritic as the value of $\tilde{t}_0$ increases. The value of the parameter $\psi$ for edges with a diameter less than 1.5 mm is almost unchanged as $\tilde{t}_0$ increases. And it shows a quick increase for edges with a diameter greater than 2.0 mm before $\tilde{t}_0$ reaches 1.0, and then changes to a slow increase. This may mean that after a sequenced process of adaption and optimization, the distributions of small vessels remain nearly unchanged for different values of $\tilde{t}_0$, but have marked affects on the main branches. 

Our work provides simulated vascular networks that exhibit a clear transition in structural pattern during the hybrid process of adaptation and optimization, changing from a `chive-like' structure to a dendritic one. Such a transition in structural pattern indicates that vascular networks in living organisms do not conform to simple global optimization described in previous studies \cite{Banavar2000,Corson2010,Katifori2010}. It appears to align with the global optimization principle in body spaces where tissues and metabolic requirements are uniformly distributed, as seen in organs like the kidney, which have regular vascular structures. Since small arteries and veins are typically abundant within certain tissue space, the vascular network at this scale may adhere to the optimization principle. Although it is not a topological phase transition, the transition of structural patterns also has some important physiological implications, as the `chive-like' structure is rarely found in the vascular systems of humans and animals. Meanwhile, variations in the parameters $\chi$ and $\psi$ reveal some of quantitative features of such structural pattern transition, particularly the growth and bifurcation of the main branches.



\section{conclusion}

In summary, we have presented a hybrid approach to describe the sequential adaptation and optimization processes of vascular networks in living organisms, such as humans and animals. The first adaptation process successfully simulates the growth of main branches in the vascular network that resemble real vascular networks, and shows its ability to deal with path dependencies of these main branches on tissues and interfaces. The subsequent optimization process demonstrates its ability to address the inefficiency of the adaptation process in optimizing small vessels, showcasing an impressive iterative efficiency. To some extent, the optimization process replaces the lengthy adaptation process after $t_0$, exhibiting only the final optimized outcomes. This hybrid approach represents a step forward from previous studies on the adaptation dynamics framework \cite{Hu2013} and the optimization framework \cite{Banavar2000,Corson2010,Katifori2010}, providing a general method to generate vascular networks that resemble real ones in living organisms. The simulated vascular networks presented in this letter exhibit several exciting features, including main branches resemble those in animal bodies and a globally optimized network that can easily control the blood pressure drop between the inlets and outlets. It shall be noted that, the ability of this hybrid approach to generate vascular networks in microcirculation is questionable, as the main factors that dominate the structure formation of vascular network in microcirculation are quite different \cite{Folkman1987,Murray1998}.

The work was supported by NSFC Grants No.12072018 and No.11602013.

\bibliography{NMAC&Circul}

\begin{thebibliography}{18}%
\makeatletter
\providecommand \@ifxundefined [1]{%
 \@ifx{#1\undefined}
}%
\providecommand \@ifnum [1]{%
 \ifnum #1\expandafter \@firstoftwo
 \else \expandafter \@secondoftwo
 \fi
}%
\providecommand \@ifx [1]{%
 \ifx #1\expandafter \@firstoftwo
 \else \expandafter \@secondoftwo
 \fi
}%
\providecommand \natexlab [1]{#1}%
\providecommand \enquote  [1]{``#1''}%
\providecommand \bibnamefont  [1]{#1}%
\providecommand \bibfnamefont [1]{#1}%
\providecommand \citenamefont [1]{#1}%
\providecommand \href@noop [0]{\@secondoftwo}%
\providecommand \href [0]{\begingroup \@sanitize@url \@href}%
\providecommand \@href[1]{\@@startlink{#1}\@@href}%
\providecommand \@@href[1]{\endgroup#1\@@endlink}%
\providecommand \@sanitize@url [0]{\catcode `\\12\catcode `\$12\catcode `\&12\catcode `\#12\catcode `\^12\catcode `\_12\catcode `\%12\relax}%
\providecommand \@@startlink[1]{}%
\providecommand \@@endlink[0]{}%
\providecommand \url  [0]{\begingroup\@sanitize@url \@url }%
\providecommand \@url [1]{\endgroup\@href {#1}{\urlprefix }}%
\providecommand \urlprefix  [0]{URL }%
\providecommand \Eprint [0]{\href }%
\providecommand \doibase [0]{https://doi.org/}%
\providecommand \selectlanguage [0]{\@gobble}%
\providecommand \bibinfo  [0]{\@secondoftwo}%
\providecommand \bibfield  [0]{\@secondoftwo}%
\providecommand \translation [1]{[#1]}%
\providecommand \BibitemOpen [0]{}%
\providecommand \bibitemStop [0]{}%
\providecommand \bibitemNoStop [0]{.\EOS\space}%
\providecommand \EOS [0]{\spacefactor3000\relax}%
\providecommand \BibitemShut  [1]{\csname bibitem#1\endcsname}%
\let\auto@bib@innerbib\@empty
\bibitem [{\citenamefont {O’Connor}\ \emph {et~al.}(2022)\citenamefont {O’Connor}, \citenamefont {Brady}, \citenamefont {Zheng}, \citenamefont {Moore},\ and\ \citenamefont {Stevens}}]{O’Connor2022}%
  \BibitemOpen
  \bibfield  {author} {\bibinfo {author} {\bibfnamefont {C.}~\bibnamefont {O’Connor}}, \bibinfo {author} {\bibfnamefont {E.}~\bibnamefont {Brady}}, \bibinfo {author} {\bibfnamefont {Y.}~\bibnamefont {Zheng}}, \bibinfo {author} {\bibfnamefont {E.}~\bibnamefont {Moore}},\ and\ \bibinfo {author} {\bibfnamefont {K.~R.}\ \bibnamefont {Stevens}},\ }\bibfield  {title} {\bibinfo {title} {Engineering the multiscale complexity of vascular networks},\ }\href {https://doi.org/10.1038/s41578-022-00447-8} {\bibfield  {journal} {\bibinfo  {journal} {Nat. Rev. Mater.}\ }\textbf {\bibinfo {volume} {7}},\ \bibinfo {pages} {702} (\bibinfo {year} {2022})}\BibitemShut {NoStop}%
\bibitem [{\citenamefont {Kinstlinger}\ \emph {et~al.}(2020)\citenamefont {Kinstlinger}, \citenamefont {Saxton}, \citenamefont {Calderon}, \citenamefont {Ruiz}, \citenamefont {Yalacki}, \citenamefont {Deme}, \citenamefont {Rosenkrantz}, \citenamefont {Louis-Rosenberg}, \citenamefont {Johansson}, \citenamefont {Janson}, \citenamefont {Sazer}, \citenamefont {Panchavati}, \citenamefont {Bissig}, \citenamefont {Stevens},\ and\ \citenamefont {Miller}}]{Kinstlinger2020}%
  \BibitemOpen
  \bibfield  {author} {\bibinfo {author} {\bibfnamefont {I.~S.}\ \bibnamefont {Kinstlinger}}, \bibinfo {author} {\bibfnamefont {S.~H.}\ \bibnamefont {Saxton}}, \bibinfo {author} {\bibfnamefont {G.~A.}\ \bibnamefont {Calderon}}, \bibinfo {author} {\bibfnamefont {K.~V.}\ \bibnamefont {Ruiz}}, \bibinfo {author} {\bibfnamefont {D.~R.}\ \bibnamefont {Yalacki}}, \bibinfo {author} {\bibfnamefont {P.~R.}\ \bibnamefont {Deme}}, \bibinfo {author} {\bibfnamefont {J.~E.}\ \bibnamefont {Rosenkrantz}}, \bibinfo {author} {\bibfnamefont {J.~D.}\ \bibnamefont {Louis-Rosenberg}}, \bibinfo {author} {\bibfnamefont {F.}~\bibnamefont {Johansson}}, \bibinfo {author} {\bibfnamefont {K.~D.}\ \bibnamefont {Janson}}, \bibinfo {author} {\bibfnamefont {D.~W.}\ \bibnamefont {Sazer}}, \bibinfo {author} {\bibfnamefont {S.~S.}\ \bibnamefont {Panchavati}}, \bibinfo {author} {\bibfnamefont {K.-D.}\ \bibnamefont {Bissig}}, \bibinfo {author} {\bibfnamefont {K.~R.}\ \bibnamefont {Stevens}},\ and\ \bibinfo {author} {\bibfnamefont {J.~S.}\ \bibnamefont {Miller}},\ }\bibfield  {title} {\bibinfo {title} {Generation of model tissues with dendritic vascular networks via sacrificial laser-sintered carbohydrate templates},\ }\href {https://doi.org/10.1038/s41551-020-0566-1} {\bibfield  {journal} {\bibinfo  {journal} {Nat. Biomed. Eng.}\ }\textbf {\bibinfo {volume} {4}},\ \bibinfo {pages} {916—932} (\bibinfo {year} {2020})}\BibitemShut {NoStop}%
\bibitem [{\citenamefont {Banavar}\ \emph {et~al.}(2000)\citenamefont {Banavar}, \citenamefont {Colaiori}, \citenamefont {Flammini}, \citenamefont {Maritan},\ and\ \citenamefont {Rinaldo}}]{Banavar2000}%
  \BibitemOpen
  \bibfield  {author} {\bibinfo {author} {\bibfnamefont {J.~R.}\ \bibnamefont {Banavar}}, \bibinfo {author} {\bibfnamefont {F.}~\bibnamefont {Colaiori}}, \bibinfo {author} {\bibfnamefont {A.}~\bibnamefont {Flammini}}, \bibinfo {author} {\bibfnamefont {A.}~\bibnamefont {Maritan}},\ and\ \bibinfo {author} {\bibfnamefont {A.}~\bibnamefont {Rinaldo}},\ }\bibfield  {title} {\bibinfo {title} {Topology of the fittest transportation network},\ }\href {https://doi.org/10.1103/PhysRevLett.84.4745} {\bibfield  {journal} {\bibinfo  {journal} {Phys. Rev. Lett.}\ }\textbf {\bibinfo {volume} {84}},\ \bibinfo {pages} {4745} (\bibinfo {year} {2000})}\BibitemShut {NoStop}%
\bibitem [{\citenamefont {Schreiner}\ and\ \citenamefont {Buxbaum}(1993)}]{Schreiner1993}%
  \BibitemOpen
  \bibfield  {author} {\bibinfo {author} {\bibfnamefont {W.}~\bibnamefont {Schreiner}}\ and\ \bibinfo {author} {\bibfnamefont {P.}~\bibnamefont {Buxbaum}},\ }\bibfield  {title} {\bibinfo {title} {Computer-optimization of vascular trees},\ }\href {https://doi.org/10.1109/10.243413} {\bibfield  {journal} {\bibinfo  {journal} {IEEE Trans. Biomed. Eng.}\ }\textbf {\bibinfo {volume} {40}},\ \bibinfo {pages} {482—491} (\bibinfo {year} {1993})}\BibitemShut {NoStop}%
\bibitem [{\citenamefont {Hu}\ and\ \citenamefont {Cai}(2013)}]{Hu2013}%
  \BibitemOpen
  \bibfield  {author} {\bibinfo {author} {\bibfnamefont {D.}~\bibnamefont {Hu}}\ and\ \bibinfo {author} {\bibfnamefont {D.}~\bibnamefont {Cai}},\ }\bibfield  {title} {\bibinfo {title} {Adaptation and optimization of biological transport networks},\ }\href {https://doi.org/10.1103/physrevlett.111.138701} {\bibfield  {journal} {\bibinfo  {journal} {Phys. Rev. Lett.}\ }\textbf {\bibinfo {volume} {111}},\ \bibinfo {pages} {138701} (\bibinfo {year} {2013})}\BibitemShut {NoStop}%
\bibitem [{\citenamefont {Corson}(2010)}]{Corson2010}%
  \BibitemOpen
  \bibfield  {author} {\bibinfo {author} {\bibfnamefont {F.}~\bibnamefont {Corson}},\ }\bibfield  {title} {\bibinfo {title} {Fluctuations and redundancy in optimal transport networks},\ }\href {https://doi.org/10.1103/physrevlett.104.048703} {\bibfield  {journal} {\bibinfo  {journal} {Phys. Rev. Lett.}\ }\textbf {\bibinfo {volume} {104}},\ \bibinfo {pages} {048703} (\bibinfo {year} {2010})}\BibitemShut {NoStop}%
\bibitem [{\citenamefont {Katifori}\ \emph {et~al.}(2010)\citenamefont {Katifori}, \citenamefont {Szöllosi},\ and\ \citenamefont {Magnasco}}]{Katifori2010}%
  \BibitemOpen
  \bibfield  {author} {\bibinfo {author} {\bibfnamefont {E.}~\bibnamefont {Katifori}}, \bibinfo {author} {\bibfnamefont {G.~J.}\ \bibnamefont {Szöllosi}},\ and\ \bibinfo {author} {\bibfnamefont {M.~O.}\ \bibnamefont {Magnasco}},\ }\bibfield  {title} {\bibinfo {title} {Damage and fluctuations induce loops in optimal transport networks},\ }\href {https://doi.org/10.1103/physrevlett.104.048704} {\bibfield  {journal} {\bibinfo  {journal} {Phys. Rev. Lett.}\ }\textbf {\bibinfo {volume} {104}},\ \bibinfo {pages} {048704} (\bibinfo {year} {2010})}\BibitemShut {NoStop}%
\bibitem [{\citenamefont {Durand}(2007)}]{Durand2007}%
  \BibitemOpen
  \bibfield  {author} {\bibinfo {author} {\bibfnamefont {M.}~\bibnamefont {Durand}},\ }\bibfield  {title} {\bibinfo {title} {Structure of optimal transport networks subject to a global constraint},\ }\href {https://doi.org/10.1103/PhysRevLett.98.088701} {\bibfield  {journal} {\bibinfo  {journal} {Phys. Rev. Lett.}\ }\textbf {\bibinfo {volume} {98}},\ \bibinfo {pages} {088701} (\bibinfo {year} {2007})}\BibitemShut {NoStop}%
\bibitem [{\citenamefont {Anderson}(1972)}]{Anderson1972}%
  \BibitemOpen
  \bibfield  {author} {\bibinfo {author} {\bibfnamefont {P.~W.}\ \bibnamefont {Anderson}},\ }\bibfield  {title} {\bibinfo {title} {More is different},\ }\href {https://doi.org/10.1126/science.177.4047.393} {\bibfield  {journal} {\bibinfo  {journal} {Science}\ }\textbf {\bibinfo {volume} {177}},\ \bibinfo {pages} {393} (\bibinfo {year} {1972})}\BibitemShut {NoStop}%
\bibitem [{\citenamefont {Chiu}\ and\ \citenamefont {Anderson}(2012)}]{Chiu2012}%
  \BibitemOpen
  \bibfield  {author} {\bibinfo {author} {\bibfnamefont {I.-S.}\ \bibnamefont {Chiu}}\ and\ \bibinfo {author} {\bibfnamefont {R.~H.}\ \bibnamefont {Anderson}},\ }\bibfield  {title} {\bibinfo {title} {Can we better understand the known variations in coronary arterial anatomy?},\ }\href {https://doi.org/10.1016/j.athoracsur.2012.05.133} {\bibfield  {journal} {\bibinfo  {journal} {Ann. Thorac. Surg.}\ }\textbf {\bibinfo {volume} {94}},\ \bibinfo {pages} {1751} (\bibinfo {year} {2012})}\BibitemShut {NoStop}%
\bibitem [{\citenamefont {Gautier}\ \emph {et~al.}(2000)\citenamefont {Gautier}, \citenamefont {Ganz}, \citenamefont {Krügel}, \citenamefont {Gill},\ and\ \citenamefont {Ganz}}]{Gautier2000}%
  \BibitemOpen
  \bibfield  {author} {\bibinfo {author} {\bibfnamefont {E.}~\bibnamefont {Gautier}}, \bibinfo {author} {\bibfnamefont {K.}~\bibnamefont {Ganz}}, \bibinfo {author} {\bibfnamefont {N.}~\bibnamefont {Krügel}}, \bibinfo {author} {\bibfnamefont {T.}~\bibnamefont {Gill}},\ and\ \bibinfo {author} {\bibfnamefont {R.}~\bibnamefont {Ganz}},\ }\bibfield  {title} {\bibinfo {title} {Anatomy of the medial femoral circumflex artery and its surgical implications},\ }\href {https://doi.org/10.1302/0301-620x.82b5.0820679} {\bibfield  {journal} {\bibinfo  {journal} {J. Bone Joint Surg. Br.}\ }\textbf {\bibinfo {volume} {82-B}},\ \bibinfo {pages} {679} (\bibinfo {year} {2000})}\BibitemShut {NoStop}%
\bibitem [{\citenamefont {Delong}(1973)}]{DELONG1973}%
  \BibitemOpen
  \bibfield  {author} {\bibinfo {author} {\bibfnamefont {W.~B.}\ \bibnamefont {Delong}},\ }\bibfield  {title} {\bibinfo {title} {Anatomy of the middle cerebral artery: The temporal branches},\ }\href {https://doi.org/10.1161/01.str.4.3.412} {\bibfield  {journal} {\bibinfo  {journal} {Stroke}\ }\textbf {\bibinfo {volume} {4}},\ \bibinfo {pages} {412} (\bibinfo {year} {1973})}\BibitemShut {NoStop}%
\bibitem [{\citenamefont {Pries}\ \emph {et~al.}(2001)\citenamefont {Pries}, \citenamefont {Reglin},\ and\ \citenamefont {Secomb}}]{Pries2001}%
  \BibitemOpen
  \bibfield  {author} {\bibinfo {author} {\bibfnamefont {A.~R.}\ \bibnamefont {Pries}}, \bibinfo {author} {\bibfnamefont {B.}~\bibnamefont {Reglin}},\ and\ \bibinfo {author} {\bibfnamefont {T.~W.}\ \bibnamefont {Secomb}},\ }\bibfield  {title} {\bibinfo {title} {Structural adaptation of microvascular networks: functional roles of adaptive responses},\ }\href {https://doi.org/10.1152/ajpheart.2001.281.3.H1015} {\bibfield  {journal} {\bibinfo  {journal} {Am. J. Physiol. Heart Circ. Physiol.}\ }\textbf {\bibinfo {volume} {281}},\ \bibinfo {pages} {H1015} (\bibinfo {year} {2001})},\ \bibinfo {note} {pMID: 11514266},\ \Eprint {https://arxiv.org/abs/https://doi.org/10.1152/ajpheart.2001.281.3.H1015} {https://doi.org/10.1152/ajpheart.2001.281.3.H1015} \BibitemShut {NoStop}%
\bibitem [{\citenamefont {Delashaw}\ and\ \citenamefont {Duling}(1988)}]{Delashaw1988}%
  \BibitemOpen
  \bibfield  {author} {\bibinfo {author} {\bibfnamefont {J.~B.}\ \bibnamefont {Delashaw}}\ and\ \bibinfo {author} {\bibfnamefont {B.~R.}\ \bibnamefont {Duling}},\ }\bibfield  {title} {\bibinfo {title} {A study of the functional elements regulating capillary perfusion in striated muscle},\ }\href {https://doi.org/https://doi.org/10.1016/0026-2862(88)90016-7} {\bibfield  {journal} {\bibinfo  {journal} {Microvasc. Res.}\ }\textbf {\bibinfo {volume} {36}},\ \bibinfo {pages} {162} (\bibinfo {year} {1988})}\BibitemShut {NoStop}%
\bibitem [{\citenamefont {Chen}\ \emph {et~al.}(2012)\citenamefont {Chen}, \citenamefont {Jiang}, \citenamefont {Li}, \citenamefont {Hu}, \citenamefont {Bu}, \citenamefont {Cai},\ and\ \citenamefont {Du}}]{Chen2012}%
  \BibitemOpen
  \bibfield  {author} {\bibinfo {author} {\bibfnamefont {Q.}~\bibnamefont {Chen}}, \bibinfo {author} {\bibfnamefont {L.}~\bibnamefont {Jiang}}, \bibinfo {author} {\bibfnamefont {C.}~\bibnamefont {Li}}, \bibinfo {author} {\bibfnamefont {D.}~\bibnamefont {Hu}}, \bibinfo {author} {\bibfnamefont {J.-w.}\ \bibnamefont {Bu}}, \bibinfo {author} {\bibfnamefont {D.}~\bibnamefont {Cai}},\ and\ \bibinfo {author} {\bibfnamefont {J.-l.}\ \bibnamefont {Du}},\ }\bibfield  {title} {\bibinfo {title} {Haemodynamics-driven developmental pruning of brain vasculature in zebrafish},\ }\href {https://doi.org/10.1371/journal.pbio.1001374} {\bibfield  {journal} {\bibinfo  {journal} {PLoS Biol.}\ }\textbf {\bibinfo {volume} {10}},\ \bibinfo {pages} {1} (\bibinfo {year} {2012})}\BibitemShut {NoStop}%
\bibitem [{\citenamefont {Mohrman}\ and\ \citenamefont {Heller}(2006)}]{Mohrman2006}%
  \BibitemOpen
  \bibfield  {author} {\bibinfo {author} {\bibfnamefont {D.}~\bibnamefont {Mohrman}}\ and\ \bibinfo {author} {\bibfnamefont {L.~J.}\ \bibnamefont {Heller}},\ }\href@noop {} {\emph {\bibinfo {title} {Cardiovascular physiology}}}\ (\bibinfo  {publisher} {McGraw-Hill Medical},\ \bibinfo {year} {2006})\BibitemShut {NoStop}%
\bibitem [{\citenamefont {Folkman}\ and\ \citenamefont {Klagsbrun}(1987)}]{Folkman1987}%
  \BibitemOpen
  \bibfield  {author} {\bibinfo {author} {\bibfnamefont {J.}~\bibnamefont {Folkman}}\ and\ \bibinfo {author} {\bibfnamefont {M.}~\bibnamefont {Klagsbrun}},\ }\bibfield  {title} {\bibinfo {title} {Angiogenic factors},\ }\href@noop {} {\bibfield  {journal} {\bibinfo  {journal} {Science}\ }\textbf {\bibinfo {volume} {235}},\ \bibinfo {pages} {442} (\bibinfo {year} {1987})}\BibitemShut {NoStop}%
\bibitem [{\citenamefont {Murray}\ \emph {et~al.}(1998)\citenamefont {Murray}, \citenamefont {Cook}, \citenamefont {Tyson},\ and\ \citenamefont {Lubkin}}]{Murray1998}%
  \BibitemOpen
  \bibfield  {author} {\bibinfo {author} {\bibfnamefont {J.}~\bibnamefont {Murray}}, \bibinfo {author} {\bibfnamefont {J.}~\bibnamefont {Cook}}, \bibinfo {author} {\bibfnamefont {R.}~\bibnamefont {Tyson}},\ and\ \bibinfo {author} {\bibfnamefont {S.}~\bibnamefont {Lubkin}},\ }\bibfield  {title} {\bibinfo {title} {Spatial pattern formation in biology: I. dermal wound healing. ii. bacterial patterns},\ }\href {https://doi.org/https://doi.org/10.1016/S0016-0032(97)00034-3} {\bibfield  {journal} {\bibinfo  {journal} {J. Franklin Inst.}\ }\textbf {\bibinfo {volume} {335}},\ \bibinfo {pages} {303} (\bibinfo {year} {1998})},\ \bibinfo {note} {part Special Issue on Biomathematics}\BibitemShut {NoStop}%
\end{thebibliography}%

\end{document}